\RequirePackage[l2tabu,orthodox]{nag} 
\documentclass[12pt,letterpaper]{article} 

\usepackage[T1]{fontenc} 
\usepackage[utf8]{inputenc} 
\usepackage{ebgaramond} 
\usepackage{tgheros} 

\usepackage[USenglish]{babel} 
\usepackage[strict,autostyle]{csquotes} 
\usepackage[babel=true]{microtype} 

\usepackage{amsmath} 
\usepackage{authblk} 
\usepackage{booktabs} 
\usepackage{caption} 
\usepackage{datetime} 
\usepackage[final]{draftwatermark} 
\usepackage{endnotes} 
\usepackage{geometry} 
\usepackage{graphicx} 
\usepackage{natbib} 
\usepackage{rotating} 
\usepackage{setspace} 
\usepackage{titlesec} 
\usepackage{url} 

\usepackage{hyperref} 
\usepackage{orcidlink} 

\newdateformat{monthyeardate}{\monthname[\THEMONTH] \THEYEAR}

\newcommand{\myname}{Geoff Boeing}

\newcommand{\myaffiliation}{Department of Urban Planning and Spatial Analysis\\University of Southern California}

\newcommand{\papertitle}{Topological Graph Simplification Solutions to the Street Intersection Miscount Problem}
\newcommand{\papercitation}{Boeing, G. 2025. \papertitle. \textit{Transactions in GIS}, published online ahead of print. \href{https://doi.org/10.1111/tgis.70037}{doi:10.1111/tgis.70037}}
\newcommand{\paperkeywords}{GIS, network science, openstreetmap, street networks, transport planning}

\graphicspath{{./figures/}}

\titleformat{\section}{\normalfont\sffamily\large\bfseries\color{black}}{\thesection.}{0.3em}{}
\titleformat{\subsection}{\normalfont\sffamily\small\bfseries\color{black}}{\thesubsection.}{0.3em}{}
\titleformat{\subsubsection}{\normalfont\sffamily\small\color{black}}{\thesubsubsection.}{0.3em}{}

\hypersetup{
    pdfauthor={\myname},
    pdftitle={\papertitle},
    pdfsubject={\papertitle},
    pdfkeywords={\paperkeywords},
    pdffitwindow=true, 
    breaklinks=true, 
    colorlinks=false, 
    pdfborder={0 0 0} 
}
\begin{document}

    \title{\papertitle\thanks{Citation info: \papercitation}}
    \author[]{\myname~\orcidlink{0000-0003-1851-6411}}  
    \affil[]{\myaffiliation}
    \date{}
    \maketitle

\begin{abstract}
Street intersection counts and densities are ubiquitous measures in transport geography and planning. However, typical street network data and typical street network analysis tools can substantially overcount them. This article explains the three main reasons why this happens and presents solutions to each. It contributes algorithms to automatically simplify spatial graphs of urban street networks---via edge simplification and node consolidation---resulting in faster parsimonious models and more accurate network measures like intersection counts and densities, street segment lengths, and node degrees. These algorithms' information compression improves downstream graph analytics' memory and runtime efficiency, boosting analytical tractability without loss of model fidelity. Finally, this article validates these algorithms and empirically assesses intersection count biases worldwide to demonstrate the problem's widespread prevalence. Without consolidation, traditional methods would overestimate the median urban area intersection count by 14\%. However, this bias varies drastically across regions, underscoring these algorithms' importance for consistent comparative empirical analyses.
\end{abstract}

\section{Introduction}

Counting is hard: despite its veneer of simplicity, counting is one of the trickiest problems in data science \citep{leibe_counting_2016, madsen_art_1999, dignazio_data_2023, elphick_how_2008}. Most real-world objects belong to fuzzy categories, resulting in subjective decisions about what to include or exclude from counts. And beyond simple definition problems, counting is plagued by challenges from sampling biases to arbitrary filtering to simple counting errors. Worse yet, miscounts' effects ripple through all subsequent analyses because counts are ubiquitous input data in nearly all statistics and models. Their initial assumptions and quality influence everything that follows and can dramatically impact our lives---for example, by allocating congressional representation or federal funding \citep{neidert_how_2025}. Yet this complexity is usually obscured by a superficial impression that counting is easy to do because its mechanics seem easy to understand. After all, everyone learns to count in kindergarten by simply enumerating the elements in a set. But counting is hard because defining that set and identifying its members are often nontrivial. Many of the world's most important analytics rely far less on flashy data science techniques than they do on counting things well and justifying those counts effectively \citep{kratz_making_2015}.

Street intersections---the junctions where two or more street segments meet---are important transport network features whose configuration contributes to urban connectivity, centrality, compactness, walkability, and safety \citep{fogliaroni_intersections_2018,sharifi_resilient_2019,cerin_determining_2022}. Despite street intersections' apparent simplicity, counting them can be nuanced and challenging in practice \citep{yang_identifying_2020,li_complex_2020}. As we will argue, this is primarily due to three modeling challenges: network nonplanarity, intersection complexity, and curve digitization. In turn, intersection miscounts can bias downstream indicators of urban form, but this bias has not been fully accounted, no easily reusable algorithms exist to mitigate it, and intersection counts often remain \textit{ad hoc} and naïve in research and practice.

This article measures this problem's extent and asks: can we algorithmically simplify a street network graph's topology to produce a more accurate and performant model? To answer this, it introduces and validates the OSMnx package's reusable algorithms that ensure that graph nodes represent individual intersections or dead-ends and that graph edges represent individual street segments. This contribution offers three main advantages over existing techniques: (1) more accurate measures including intersection counts/densities, street segment lengths, node degrees, and betweenness centralities; (2) a topologically-corrected model that better represents the real world; and (3) information compression (i.e., fewer nodes/edges than in the original model) that can drastically speed-up subsequent network algorithms that scale with node or edge count.

The rest of this article is organized as follows. First it explains why street intersection miscounts happen and why they matter. Then it introduces the OSMnx street network modeling package's method of building a graph model and performing algorithmic topological edge simplification and node consolidation. These algorithms automatically generate the best possible model to match transport planning and urban design theory---and in turn yield more accurate counts. Finally it validates these methods and analyzes their outputs (i.e., intersection counts) to compare how this foundational measure can vary drastically if using less-sophisticated modeling techniques.

\section{Causes and Effects of Intersection Miscounts}

\subsection{Street Network Models}

We begin with some preliminary theory necessary to explain the problem and solutions. Although these concepts, problems, and solutions generalize to any nonplanar spatial network, we focus on urban street networks here. We define a \textit{street} as a thoroughfare in an urban built environment. An \textit{intersection} exists wherever two or more streets connect with each other and a \textit{dead-end} is the terminus of a street. A street \textit{segment} is the portion of a street lying between a pair of intersections and/or dead-ends.

Street networks are typically modeled as spatial \textit{graphs} \citep{barthelemy_spatial_2011,marshall_street_2018}. These mathematical abstractions consist of a set of elements called \textit{nodes} (e.g., intersections and dead-ends) connected to each other by a set of links called \textit{edges} (e.g., street segments). A \textit{multigraph} can have multiple (called \textit{parallel}) edges between a single node pair \citep{newman_networks:_2010}. A graph can contain one or more connected \textit{components}---disjoint sets of nodes each forming their own subgraphs \citep{barthelemy_spatial_2022}. Edges can be directed or undirected. Two nodes are \textit{neighbors} if they are linked by an edge, but more specifically, if a directed edge links from node \textit{u} to node \textit{v}, then \textit{v} is a \textit{successor} of \textit{u}. An edge is \textit{incident} to the nodes it connects, and a node's \textit{degree} measures how many edges connect to it \citep{trudeau_introduction_1994}. If we have a directed graph, $G$, comprising a set of nodes, $N$, and a set of edges, $E$, such that:

\begin{equation}
    \label{eq:example_graph}
    G = (N, E)
\end{equation}
\begin{equation}
    \label{eq:example_nodes}
    \{u, v\} \subseteq N
\end{equation}
\begin{equation}
    \label{eq:example_edges}
    (u, v) \in E
\end{equation}

then we can say for $G$ that: (1) edge $(u, v)$ links from node $u$ to node $v$, (2) edge $(u, v)$ is incident to $u$ and to $v$, and (3) $v$ is a successor of $u$.

Graphs can model both street network \textit{geometry} and \textit{topology} \citep{fischer_spatial_2014}. Geometry refers to spatial position and shape, and includes lengths, areas, distances, and angles. Topology refers to elements' relative configuration and connections \citep{lin_complex_2013}. A \textit{nonplanar} graph's topology cannot be represented in a plane such that its edges cross only at nodes: most real-world urban street networks are nonplanar \citep{eppstein_studying_2008,boeing_planarity_2020}. Modeling both geometry and topology are important because street networks are defined, characterized, and constrained by both \citep{fischer_spatial_2014}.

Common data sources for urban street network modeling include the U.S. Census Bureau's TIGER/Line and OpenStreetMap \citep{zielstra_assessing_2013}. TIGER/Line provides street centerline geometries across the U.S. in ESRI Shapefile format \citep{esri_shapefile_1998}. Shapefiles and data representations that adhere to the Simple Features specification can fully represent network geometry but cannot fully represent nonplanar topology. OpenStreetMap is a worldwide mapping platform and geospatial database that anyone can contribute to \citep{jokar_arsanjani_openstreetmap_2015}. Unlike TIGER/Line Shapefiles, OpenStreetMap data provide both geometry and topology for streets worldwide. Its data are generally high quality, with some exceptions, notably in China and sub-Saharan Africa \citep{barrington-leigh_worlds_2017,basiri_quality_2016,barron_comprehensive_2014,corcoran_analysing_2013,neis_street_2011,haklay_how_2010,girres_quality_2010,baumann_assessing_2020,grinberger_analysis_2021}.

\subsection{Three Causes of Intersection Miscounts}

Streets and intersections are functionally two-dimensional, but graph models represent them as one-dimensional lines and dimensionless points, respectively. This parsimony lets us apply the mathematical tools of graph theory and linear algebra to solve problems using these models, but it also comes with drawbacks. Street intersections, particularly the complex kind common in modern car-centric urban areas, are fuzzy objects for which most data sources do not provide a simple 1:1 representation. This results in what \citet{fisher_models_1999} identifies as spatial uncertainty attributable to vagueness and ambiguity.

\begin{figure*}[tb]
    \centering
    \includegraphics[width=1\textwidth]{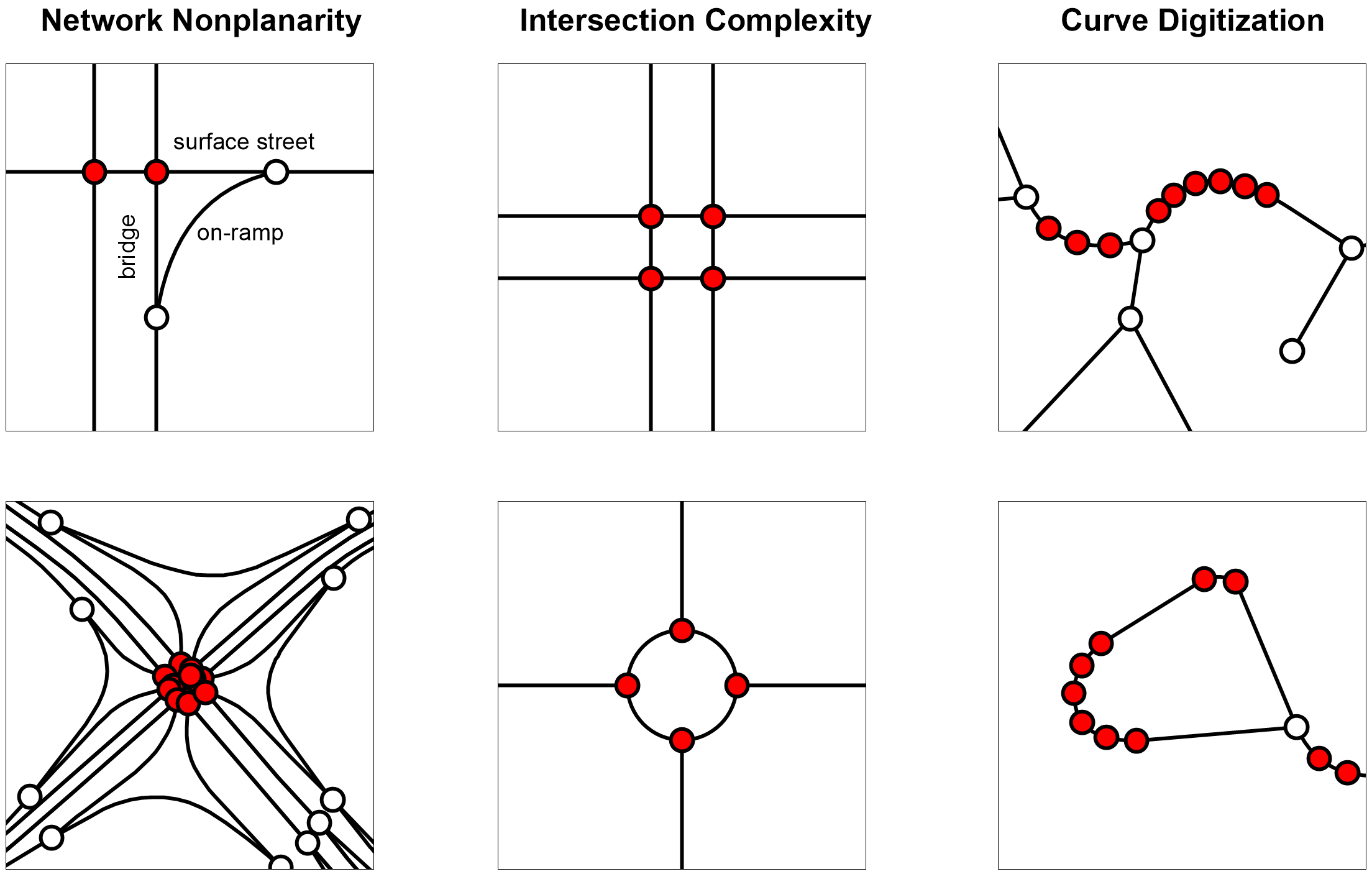}
    \caption{Three causes of intersection miscounts. Lines represent edges, white circles represent true nodes (i.e., intersections/dead-ends), red circles represent false nodes at planar line intersections, redundant/duplicative intersections, or interstitial curvilinear geometry vertices.}\label{fig:fig_miscount_causes}
\end{figure*}

Analysts often use desktop GIS to calculate intersection counts from street geometry source data, such as TIGER/Line Shapefiles or OpenStreetMap extracts. Such data represent street geometries as centerlines, so intersections could be considered either as these planar line intersections (if only geometric data are available, such as from TIGER/Line) or topological intersections (if topological data are available, such as from OpenStreetMap). But such data sources can exhibit substantial bias due to three main causes illustrated by Figure~\ref{fig:fig_miscount_causes}: network nonplanarity, intersection complexity, and curve digitization.

\subsubsection{Network Nonplanarity}

The nonplanarity problem arises when using purely geometric data \citep{boeing_planarity_2020}. For example, the traditional GIS technique for counting intersections uses two-dimensional street network geometry data, such as TIGER/Line Shapefiles of street centerlines, then intersects those line geometries and counts the intersection points. However, street centerlines also intersect in two dimensions at overpasses and underpasses, as seen in Figure~\ref{fig:fig_miscount_causes} (left).

This planar/geometric counting method is obsolete (though still in use) today given that better data sources, algorithms, and toolkits are available: nonplanar intersection counts offer an improvement by using topological street network data, often from OpenStreetMap. In other words, the relationships between the model's elements are formally topological and are preserved under continuous deformations and transformations of the underlying space. Rather than counting planar centerline intersections as street intersections, a nonplanar counting technique instead only counts true intersections in the three-dimensional street network topology \citep{boeing_planarity_2020}. That is, if the network is modeled as a nonplanar graph of street centerlines (edges) and intersections/dead-ends (nodes), then the street intersections would be the non-dead-end nodes (i.e., the subset of nodes with degree $> 1$).

\subsubsection{Intersection Complexity}

Although nonplanar node counts offer an improvement over geometric centerline intersection counts, they are only accurate insofar as there exists a 1:1 correspondence between real-world intersections and graph nodes. Unfortunately this is almost never the case: in the real world, intersections are often complex fuzzy objects. A few examples illustrate this. For instance, most street network data---either geometric or topological---represent a single divided road with a median as a pair of side-by-side one-way centerline edges in reciprocal directions. The intersection of such divided roads produces four nodes (i.e., $2 \times 2$ centerlines) in models deriving from these data \citep{barrington-leigh_global_2020}. Thus, using modern data sources and standard modeling techniques, an analyst would count four intersections at the single complex intersection of two divided roads, as seen in Figure~\ref{fig:fig_miscount_causes} (top center).

As another example, a single four-way intersection consisting of a single roundabout will often be represented as a circumferential line geometry with four streets intersecting it, producing four three-way intersections in the data and subsequent models, as seen in Figure~\ref{fig:fig_miscount_causes} (bottom center). Other features such as slip lanes and certain turn lane digitization similarly create complex geometries and topologies that are important for faithfully representing the street network, but produce multiple graph nodes for what is considered a single intersection in the real world. Worst of all, the nature of such bias means that individual neighborhoods with the most car-centric complex intersections are the most over-represented in subsequent intersection density analyses.

\subsubsection{Curve Digitization}

Finally, some data sources represent curvilinear street centerline geometries' vertices as nodes. OpenStreetMap is a prominent example of this: its data model uses the same \enquote{node} data type for curving line geometry vertices in addition to representing street intersections and dead-ends. Thus, a single curving street comprises a set of many tiny straight-line segments, as seen in Figure~\ref{fig:fig_miscount_causes} (right). There is no \textit{prima facie} differentiation for these different kinds of nodes in its data model, so analysts cannot simply count nodes as a proxy for intersections.

\subsection{The Problem of Miscounts}

What gets measured gets done \citep{giles-corti_creating_2022}. Intersection counts are ubiquitous in transport research and practice as foundational input data in innumerable analyses, models, and reports \citep{ewing_travel_2010}. Notably, due to its simplicity and interpretability, intersection density (i.e., count normalized by area) is the most common measure of compact and sustainable street network design in practice \citep[e.g.,][]{congress_for_the_new_urbanism_street_2021,duany_smart_2010,marshall_street_2010,boeing_off_2021,liu_generalized_2022,higgs_global_2024}.

Yet, too often, intersections are miscounted---potentially grossly so---by transport scholars and practitioners due to the three aforementioned problems: network nonplanarity, intersection complexity, and curve digitization. These problems plague all spatial network software equally because they are source data problems arising from the underlying limitations of data formats and standards and the conceptual complexity of fuzzy sets. Working in a small study site, an analyst could mitigate these biases with meticulous hand-counting. But this mitigation strategy becomes intractable when analyzing large cities, metropolitan regions, or cross sections of many such study sites.

Intersection counting has thus emerged as a pressing transport research methods problem \citep{li_complex_2020,li_polygon-based_2014,wu_unsupervised_2022,yang_detecting_2022}. Different research groups have attempted clustering methods \citep{mackaness_automating_1999,touya_road_2010,zhou_experimental_2015}, deep learning methods \citep{li_complex_2020,li_automatic_2019}, and topological methods \citep{yang_identifying_2020,yang_detecting_2022,zhang_road_2005}. However, these solutions emphasize correcting counts rather than correcting the underlying model misrepresentation itself, tend to focus only on complex highway interchanges, and require advanced technical skills or intensive preprocessing labor to re-implement. Worse, the extent of the miscount problem itself has not been quantified and no easy-to-use tools exist for researchers or practitioners to mitigate it.

To correctly measure street network characteristics like intersection densities, street segment lengths lengths, and node degrees, we need accurate models for intersection counts as foundational input data. Algorithms to simplify a graph's topology from the raw source data to accurately model individual intersections/dead-ends as individual nodes would provide a more faithful representation and offer better measures. They would also offer better downstream performance because consolidating edge fragments and complex intersections would result in a more parsimonious model, and many network algorithms' performance scales with node/edge count. In other words, these \enquote{simplified} models would be both more accurate and more performant.

\section{Algorithms and Analysis}

To address this need, we introduce the open-source algorithms\endnote{The source code is freely available at https://github.com/gboeing/osmnx} for topological edge simplification and node consolidation in OSMnx, a Python package\endnote{Installation instructions and usage guide are available at https://osmnx.readthedocs.io/} for modeling and analyzing street networks and any other geospatial data from OpenStreetMap \citep{boeing_modeling_2025}. Then we explain how we validate and analyze these algorithms' results. OSMnx models urban street networks as nonplanar directed multigraphs to solve the aforementioned network nonplanarity problem, and it uses these edge simplification and node consolidation algorithms to solve the curve digitization problem and attenuate the intersection complexity problem, respectively.

\begin{figure*}[tbp]
    \centering
    \includegraphics[width=1\textwidth]{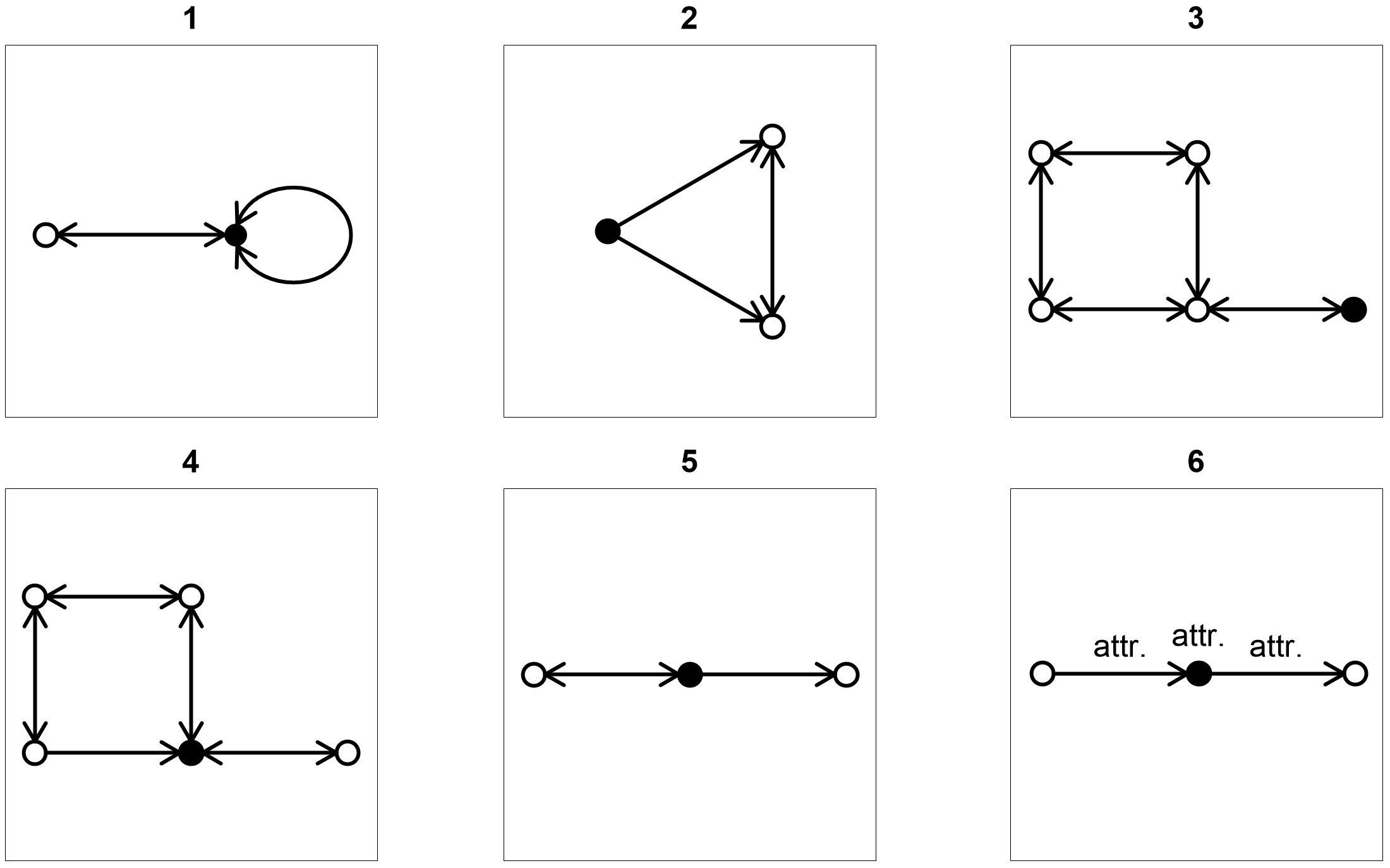}
    \caption{The edge simplification algorithm's six rules to identify a true endpoint node (i.e., an intersection or dead-end). Black circles represent the node in question, white circles represent other nodes, and arrows represent directed edges.}\label{fig:fig_edge_simp_algorithm}
\end{figure*}

\subsection{Edge Simplification Algorithm}

OSMnx's edge simplification algorithm solves the curve digitization problem, such as occurs with OpenStreetMap raw data as discussed earlier. It simplifies the graph's topology by removing all nodes that are not \enquote{true} edge endpoints (i.e., intersections or dead-ends), then creating new (simplified) edges directly between those true endpoints and retaining the full geometries of the original edges as attributes on those new edges. The full algorithm can be found in the package (see endnotes), but we summarize it here.

First the algorithm generates the set of all true endpoint nodes in the graph. This set comprises all nodes that satisfy at least one of the following six rules, illustrated by Figure~\ref{fig:fig_edge_simp_algorithm}.

\begin{enumerate}
    \item The node is its own neighbor---that is, it has an incident self-loop edge.
    \item The node's incident edges are all incoming or all outgoing.
    \item The node has exactly 1 neighbor (in which case it is a dead-end).
    \item The node has >2 neighbors (in which case it is an intersection of >2 streets).
    \item The node has 2 neighbors but degree 3 (indicating a transition from one-way to bidirectional travel).
    \item If configured, either (a) the node possesses a user-defined attribute explicitly denoting it as an endpoint, or (b) the node's incident edges possess a user-defined attribute and have different values than each other in that attribute.
\end{enumerate}

\begin{figure*}[b!]
    \centering
    \includegraphics[width=1\textwidth]{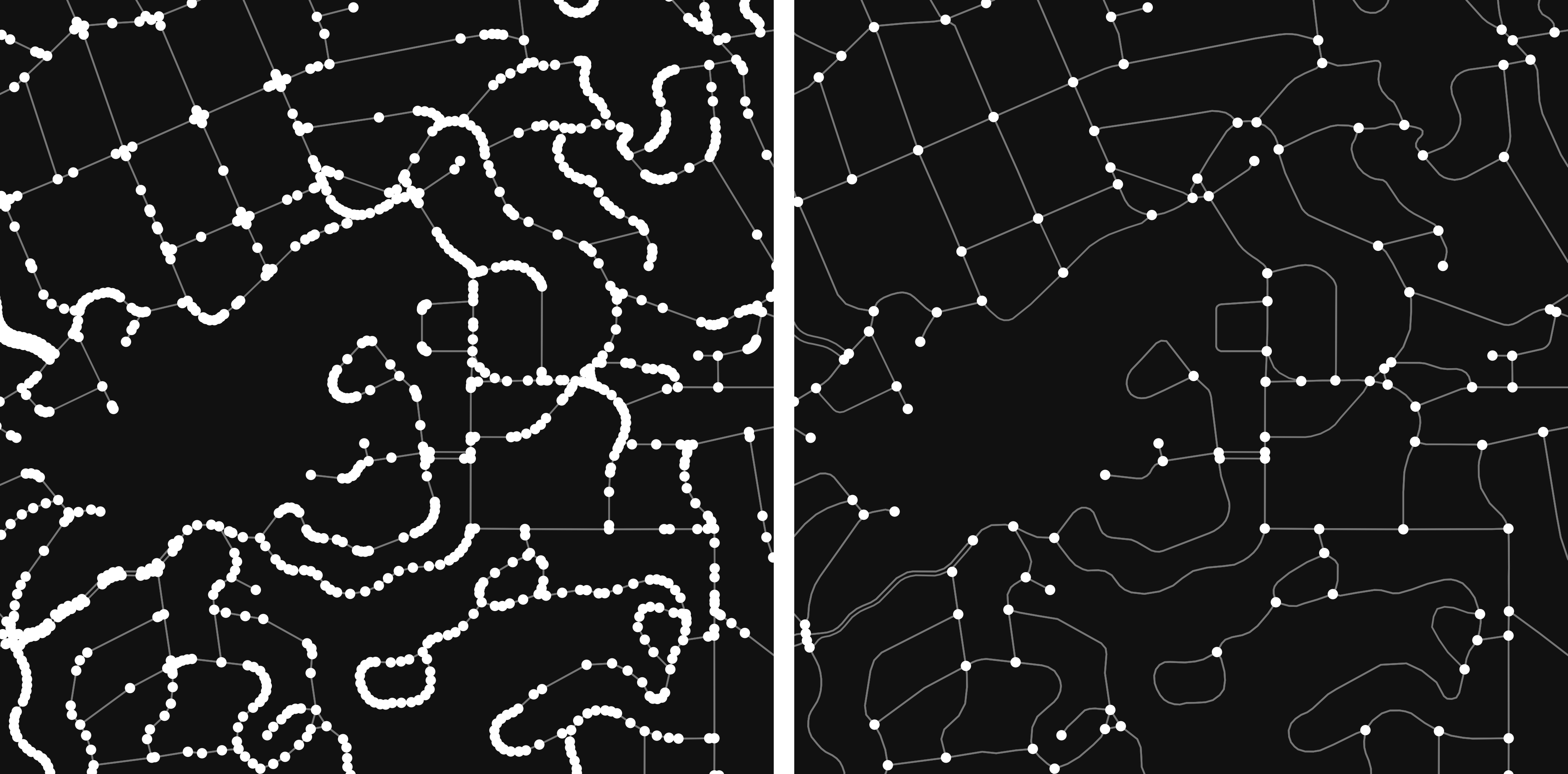}
    \caption{Before (left) and after (right) running the edge simplification algorithm on an urban street network's graph. Circles represent nodes and lines represent edges.}\label{fig:edge_simplification}
\end{figure*}

The first five rules represent the strict mode of endpoint determination by rigorously defining what belongs to the category of \enquote{intersection or dead-end.} The last rule is user-configurable to relax this strictness. This offers the user nearly infinite flexibility in parameterizing the algorithm to obtain a precise desired result. In fact, any node can be retained by parameterizing the last rule as desired. For example, nodes at \enquote{elbow bend} two-way intersections can be retained if desired when the street name or ID changes at the elbow.

Next, the algorithm generates the set of all paths to be simplified between these true endpoint nodes. For each endpoint node \textit{u}, the algorithm follows the path(s) from \textit{u} through each its successor nodes, to each of these successors' successor nodes, and so on iteratively until it reaches the next endpoint node \textit{v} along each path. Nodes \textit{u} and \textit{v} are \enquote{true} neighbors, and each node in the path between them is an \enquote{interstitial} node that is just a curving centerline's geometric vertex. Finally, for each path to simplify, the algorithm creates a new graph edge from \textit{u} to \textit{v} and adds a \enquote{geometry} attribute to it, representing the original curving line geometry from \textit{u} to \textit{v} through the nodes representing interstitial line vertices. Then it removes all those nodes representing interstitial line vertices from the graph.

As Figure~\ref{fig:edge_simplification} demonstrates, the final result is a graph with a much simpler topology---and far fewer nodes and edges---than the original graph, yet it offers a more accurate model of the street network's real-world features (i.e., intersections, dead-ends, and street segments). In particular, this enables more accurate measurement of street segment lengths between intersections. This algorithm can only be run once (because the information compression is lossy) and should only be run once (because there is no benefit from additional runs, since it can be parameterized upfront for infinitely customizable precise results from a single run).

\begin{figure*}[htbp]
    \centering
    \includegraphics[width=0.79\textwidth]{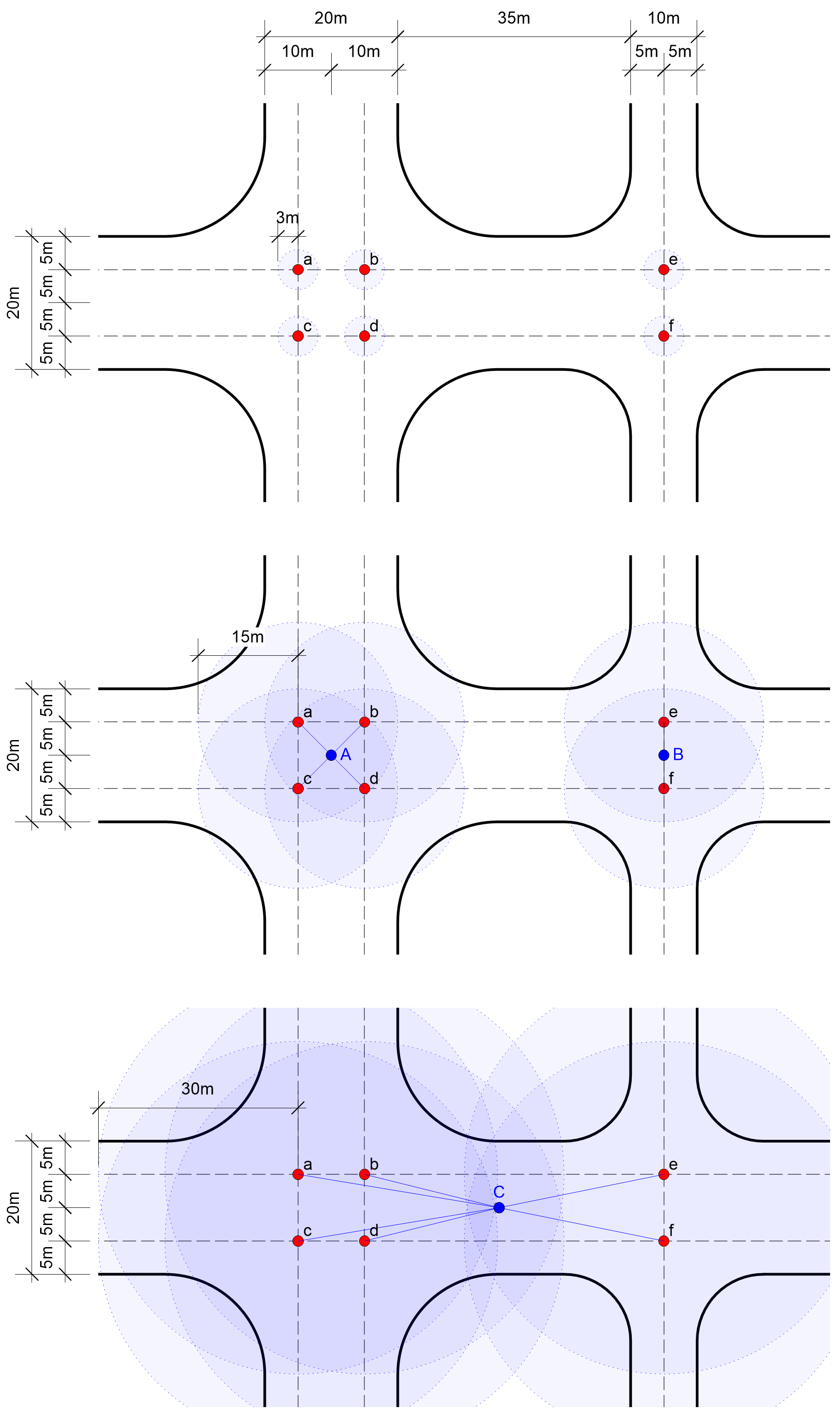}
    \caption{Node consolidation algorithm with three different tolerance parameterizations. Red circles/lines are nodes/edges before consolidation, blue are after. Top: too conservative and nodes do not merge (false negative). Middle: appropriate (true positive). Bottom: too aggressive and two separate intersections merge together (false positive).}\label{fig:node_consolidation}
\end{figure*}

\subsection{Node Consolidation Algorithms}

After edge simplification is performed, OSMnx's node consolidation algorithms attenuate the intersection complexity problem. There are two variants: geometric consolidation and topological consolidation.

\subsubsection{Geometric Node Consolidation}

The OSMnx geometric node consolidation algorithm works as follows. First, it geometrically buffers the set of graph nodes with degree $> 1$ by some user-defined tolerance (i.e., buffer radius).  As illustrated by Figure~\ref{fig:node_consolidation}, the user may specify a constant tolerance (e.g., 10 meters) to use consistently across the graph, or individual per-node tolerances for context sensitivity and precision. The latter parameterization offers the user near-infinite flexibility, so the scale and detail of the local network and its digitization are important to consider. Ideally the user would parameterize this according to the irregular urban design standards within the study site---e.g., smaller in fine-grained networks, larger in coarser-grained ones, with caution paid to the ubiquitous bias-variance trade-off. Next the algorithm merges overlapping node buffers into unified geometries and counts those geometries. Thus a single complex intersection's multiple nodes are merged into a single meta-geometry representing it as a unified whole.

\subsubsection{Topological Node Consolidation}

While geometric consolidation yields more accurate intersection counts, they may still be flawed. Geometric consolidation ignores the true topology of the network: things that lie near each other in two dimensions may be topologically distant from each other when constrained to a spatial graph embedded in three dimensions. This phenomenon is relatively common in urban street networks. As one example, consider an intersection of surface streets that lies beneath the intersection of a highway overpass and on-ramp. Although the planar Euclidean distance between them is approximately 0, these two intersections are, in reality, distinct junctions in the three-dimensional street network and should not be merged together into a single consolidated node. Another example illustrates this in a residential neighborhood. Traffic engineers sometimes use street bollards to divert four-way intersections into two separate two-way intersections, constraining neighborhood traffic flow to turns only. In such cases we should not merge these two intersections. They are topologically distinct and disconnected in the real world, and thus should not be modeled as a single four-way intersection where such does not exist.

\begin{figure*}[tbp]
    \centering
    \includegraphics[width=1\textwidth]{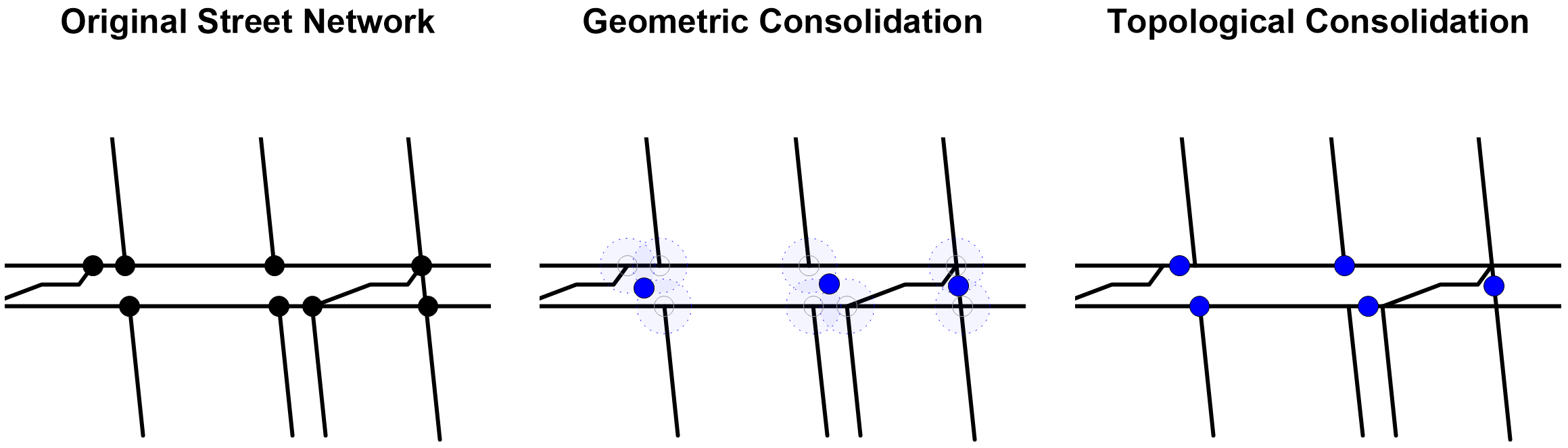}
    \caption{Spatially proximate but topologically remote nodes during consolidation. Left: original network. Center: geometric consolidation incorrectly merges unconnected nodes within spatial buffer distance. Right: topological consolidation correctly merges nodes only when they are connected within buffer distance along the network.}\label{fig:fig_topo_node_consolidation}
\end{figure*}

Instead, we need to attenuate the complex intersection problem without accidentally consolidating these \textit{spatially proximate} but \textit{topologically remote} nodes into a single node, as seen in Figure~\ref{fig:fig_topo_node_consolidation}. To address this, the OSMnx topological node consolidation algorithm consolidates intersections topologically rather than geometrically by merging nodes within some network (rather than Euclidean) distance of each other. Calculating a complete network distance matrix is computationally expensive, so this algorithm exploits a fast heuristic: it merges nodes within some tolerance geometrically (as above) into clusters, induces a subgraph on each cluster's set of nodes, then divides this subgraph into its weakly connected components. These components are the sets of nodes that are topologically connected to each other within that tolerance. Finally---but importantly---this algorithm topologically rebuilds the graph model after consolidating complex intersections, by reconnecting \enquote{broken} edges to the new consolidated nodes. This results in more accurate node degree measures downstream.

This technique offers several benefits. First, such topologically-consolidated intersection counts provide a more accurate solution to the problem of counting intersections. Accordingly, in many situations we should expect lower (and more accurate) intersection counts from this method than from even the geometrically-consolidated intersection counts, given the examples of spatial proximity but topological remoteness discussed above. Second, by reconnecting edges it properly reconstructs the street graph, producing accurate node degrees and a routeable model that better represents the real world---yet with fewer nodes. This means faster downstream graph algorithms as memory usage and time complexity generally scale with the number of nodes in the graph.

Node consolidation can and should only be run once, for the same reasons discussed earlier for the edge simplification algorithm. Additionally, edge simplification should be run prior to node consolidation, to remove interstitial nodes before consolidating those that remain.

\subsection{Assessment and Validation}

To assess the global extent of the intersection miscount problem, we model the street networks of every urban area in the world then run the edge simplification and node consolidation algorithms. The urban area boundaries come from the Global Human Settlement Layer's Urban Centres Database \citep{florczyk_description_2019}. We use OSMnx to model the drivable street network within each of these 8,910 boundaries, then perform edge simplification in standard strict mode, and then perform topological node consolidation \citep{boeing_street_2022}. For scalability in this study, we parameterize the latter algorithm with a per-urban-area constant tolerance, drawn from the Atlas of Urban Expansion's empirical average road width data set \citep{angel_et_al_atlas_2016}, by matching each urban area to its nearest neighbor in the road width data set. This configuration thus varies between urban areas but imperfectly assumes a constant parameter within each urban area. However, it offers a good initial first-order approximation assessment that allows for global scalability and follows the premise that proximate cities are more likely to share similar urbanization trends and thus more comparable street design. Finally, we count the intersections (nodes with degree $> 1$) before and after graph simplification.

To validate these algorithms, we then employ a manual qualitative case study process across a random sample of 32 world urban areas stratified by eight geographical regions (from the Atlas of Urban Expansion), high versus low population (using OECD definitions), and high versus low intersection density (using original calculated values). To validate the edge simplification algorithm, we compare these urban areas' graphs, by hand, to the original OpenStreetMap raw data and satellite imagery to verify that all graph nodes correspond to true intersections or dead-ends, and vice versa. This algorithm is deterministic and expected to perform with 100\% accuracy.

To validate the (topological) node consolidation algorithm, for each of these sampled urban areas we randomly sample 120 graph nodes, divided evenly between consolidated and not-consolidated nodes. Each node's consolidation (or not) result is then validated by qualitatively cross-referencing it, by hand, against Google Street View and satellite imagery data. In comparing these resources, the validator asks whether the node(s) in question would be considered a single intersection by a pedestrian or motorist at that intersection. There are four categories of node consolidation validation result: (1) \enquote{true positive,} where the node should be consolidated and is, (2) \enquote{true negative,} where the node should not be consolidated and is not, (3) \enquote{false positive,} where the node should not be consolidated but is, and (4) \enquote{false negative,} where the node should be consolidated but is not. In keeping with the spirit of fuzzy sets, within the \enquote{true positive} category we create a subcategory for \enquote{true positive but insufficiently robust.} For example, if a sampled validation node is in a roundabout comprising five nodes that should have been consolidated into one, but the algorithm only consolidated four of those five, then this sampled node is assigned this result subcategory. In other words, this result improves on the original model and is mostly correct, but incomplete.

\section{Validation and Assessment Results}

The initial set of 8,910 urban area street network models collectively comprise over 160 million nodes and 320 million edges before the simplification algorithms are applied. As noted earlier, the raw data from OpenStreetMap uses the same \enquote{node} objects to represent both intersections/dead-ends and curving centerlines' geometric vertices. Thus these initial models are substantially inflated with unnecessary interstitial graph nodes. Specifically, after running the edge simplification algorithm, these models collectively comprise just 37 million nodes and 53 million edges. From an information compression perspective, this represents approximately a 77\% reduction for the nodes and 83\% for the edges. This is lossy compression yet it improves model fidelity: 77\% of the original nodes were interstitial and thus topologically unnecessary in the sense of the aforementioned curve digitization problem. The manual validation reveals that the edge simplification algorithm performs with 100\% accuracy. This is unsurprising as it is entirely deterministic: we expect such a result if the algorithm's implementation is merely bug-free.

These edge-simplified graphs then serve as inputs into the node consolidation algorithm, which reduces the median urban area intersection count by a further 12\%, from 978 to 857 intersections (i.e., 121 of them were redundant in the sense of the aforementioned intersection complexity problem). To put these numbers another way, without consolidation, traditional methods would overestimate the median urban area intersection count by 14\%. However, there is geographical heterogeneity. The median urban area in Eastern Europe sees a 6\% reduction in intersection count after consolidation, whereas the median in Western Europe sees 13\%. Some urban areas see relatively small reductions, such as Dhaka (7\%) and Seoul (8\%), whereas others see much larger reductions, such as Sydney and Singapore (26\% each). This variation is largely a function of the local digitization of the street network: cities with more fine detail (such as individual turn lanes, slip lanes, complex intersection and roundabouts, etc.) digitized have more initial nodes to be consolidated into a single true intersection. This bias is more prevalent in wealthier parts of the world, possibly due to greater community interest and resources for digital mapping. Using corresponding data from the Global Human Settlement Layer, we find that urban areas' (log) reduction in intersection count after consolidation correlates significantly with both  (log) gross domestic product based on purchasing power parity (Pearson correlation coefficient $r=0.25$, $p<0.001$) and (log) nighttime light emission ($r=0.38$, $p<0.001$).

Now the question becomes: are these reductions in node count legitimate, or merely false positives being incorrectly consolidated? To that end, the node consolidation algorithm's validation can be considered from two perspectives, (1) the validity of the algorithm itself, and (2) the validity of its parameterization here. From the first perspective, this deterministic algorithm unsurprisingly performs with 100\% accuracy---that is, each sampled node's result aligns with expectations given this parameterization. From the second perspective, the algorithm performs imperfectly, but better than expected given this simplified constant parameterization per urban area used for global scalability in this preliminary demonstration.

Among nodes that were consolidated (i.e., the \enquote{positives}), the validation reveals a 98\% true positive rate and a 2\% false positive rate. Among these true positives, around 5\% were \enquote{insufficiently robust,} meaning the algorithm was correct in consolidating a complex intersection but failed to fully merge all the nodes that the validator deemed integral during their qualitative assessment. Among nodes that were not consolidated (i.e., the \enquote{negatives}), the validation reveals an 87\% true negative rate and a 13\% false negative rate. In other words, our simplified parameterization here was likely overly conservative, leading to more false negatives than false positives through what is essentially algorithmic underfitting. This suggests that the extent of the intersection miscount problem could be even greater than our conservative assessment reveals.

\section{Toward Better Models and Measures}

Counting seems simple because its mathematics are simple. But in reality it can be difficult to define what to count and how to count it. Street network intersections exemplify this challenge. In transport planning, innumerable models, measures, and standards---from sustainable development certifications to resilience simulations---rely on intersection counts as input data. Accurate intersection densities, node degrees, and street segment lengths all depend on accurate street network models. Yet modeled intersections can be surprisingly difficult to count accurately, resulting in biased measures that propagate downstream. This study identified the sources of these biases, measured their prevalence, and presented and validated algorithms to better model street intersections for better data-driven, evidence-informed transport planning. These algorithms are generalizable but their implementation here is tied to one data source (OpenStreetMap) or others matching its structure. These open-source, freely reusable methods produce network models with a closer 1:1 relationship between graph nodes and real-world intersections/dead-ends.

Specifically, these methods address the three main causes of the intersection miscount problem: network nonplanarity, curve digitization, and intersection complexity. OSMnx resolves the nonplanarity problem by modeling networks as nonplanar directed multigraphs. It resolves the curve digitization problem through its edge simplification algorithm. And it attenuates the intersection complexity problem with its node consolidation algorithms.

Addressing this intersection miscount problem is essential. \citet{boeing_planarity_2020} quantified the significant extent of the nonplanarity problem plaguing the literature. The present study extends this assessment to the curve digitization and intersection complexity problems. The former does not afflict all data sources, but must be addressed when working with ubiquitous data sources like OpenStreetMap where we find that some 77\% of urban street network nodes worldwide are merely interstitial geometric line vertices. However, essentially all data sources suffer from the intersection complexity problem due to divided roads, slip lanes, roundabouts, interchanges, complex turning lanes, etc. If unaddressed, our assessment shows that typical intersection counts (and downstream densities) would be overestimated by 14\%.

This bias's heterogeneity particularly hinders comparative urban analytics. If a researcher were measuring street network density or connectedness for, say, a walkability audit or a resilience study, this bias would significantly affect their results' interpretation. For example, without first running the node consolidation algorithm, a standard technique like counting the nonplanar non-dead-end nodes in Sydney would result in a 26\% overestimate of true intersections, but only an 8\% overestimate in Seoul. This would not only misrepresent network density, but unevenly so across study sites. This in turn unevenly affects downstream monitoring of progress toward urban sustainability goals and the ability of planners to target interventions in a biased information landscape.

Our conservative parameterization here is not perfect, but it does mean that we did not overshoot on our estimates of the intersection miscount problem's extent. That is, our estimates are conservative due to the underfitting that minimizes false positives, as shown by the validation's 98\% true positive rate. The validation also suggests that if parameterized perfectly, these algorithms would perform with perfect precision. This is trivial for the edge simplification algorithm, but, as we saw, the node consolidation algorithm's simplified constant parameterization used here for this preliminary demonstration resulted in underfitting (high bias and low variance). However, as mentioned earlier, OSMnx allows users to specify per-node consolidation tolerances. If well-defined, these values would by definition produce accurate results, but are potentially labor intensive and risk overfitting in the sense that such a parameterization is not generalizable to other places (low bias and high variance).

In addition to these model accuracy benefits, these methods also improve downstream network algorithms' computational tractability and runtime. This has significant implications in practice. For example, our assessment shows that an algorithm with quadratic time complexity $O(n^2)$ would complete roughly 20 times faster after edge simplification, on average. The two algorithms together combine to yield nearly 80\% information compression across the graph nodes globally. Best of all, these parsimonious graphs are also more faithful models of the real world.

The advantages and shortcomings discussed in this article open the door to future research. From a methodological perspective, estimating appropriate per-node tolerances for node consolidation offers an important avenue for further work. As discussed in the background section, some work has been done to this end, but better algorithms to identify which nodes constitute complex intersections and how to merge them would improve the underfitting seen here while obviating the labor required to manually define per-node tolerances. From an empirical perspective, our findings suggest that much work needs to be done to better measure street network intersection counts, densities, and degrees to support practitioners with accurate foundational inputs. This is a key opportunity for future large-scale studies to explore the extent to which intersection misrepresentation affects various downstream spatial analyses.

\section{Conclusion}

Compact and connected street networks are central to transport planning goals of active travel, emissions reduction, sustainability, and resilience. Achieving these goals requires accurate models and measurement methods. But the longstanding rough-approximation methods of traditional intersection counts using typical data sources introduce bias and preclude a true accounting of the transport system's characteristics. This article measured the extent of this problem and presented reusable methods from the OSMnx package to mitigate it.

We argue that transport planners cannot expect accurate intersection counts from traditional methods and common data sources. Worse, these biases vary across cities and regions, thwarting consistent comparative urban analyses. To address this, our contribution is threefold: (1) topologically-corrected parsimonious models that better represent the real world; (2) more accurate street network measures---including intersection counts/densities, node degrees, and street segment lengths---for transport planning; (3) information compression to improve downstream graph algorithm memory use and runtime, improving analytical tractability. In an era of rapid urbanization, transport planners need good spatial models for evidence-informed planning. This article takes a step in that direction toward better models for better empirical science.

\section*{Acknowledgments}

This work was funded in part by a grant from the California Department of Transportation and the Pacific Southwest Region University Transportation Center. The author wishes to thank Yuquan Zhou, Jiyoon Kim, and Jaehyun Ha for assistance with validation, assessment, and visualization.

\section*{Declarations}

The author has no conflicts of interest to declare.

\section*{Data and Code Availability Statement}

The end notes contain further details on the source code repository and package documentation. The source data used in this study is publicly available from \citet{angel_et_al_atlas_2016}, \citet{florczyk_description_2019}, and https://www.openstreetmap.org. The source code of the OSMnx \texttt{simplification} module version 2.0.0, which implements the algorithms described in this article, is available at https://github.com/gboeing/osmnx.

\IfFileExists{\jobname.ent}{\theendnotes}{}

\setlength{\bibsep}{0.00cm plus 0.05cm} 
\bibliographystyle{apalike}
\bibliography{references}

\end{document}